\documentclass[12pt]{article}
\usepackage[dvips]{graphicx}
\usepackage[usenames]{color}
\usepackage{latexsym}
\def\leq{\underline{<}}
\begin{document}
\title{
Self-organizing social hierarchies in a timid society
}
\author{Takashi Odagaki and Masaru Tsujiguchi\\
 Department of Physics, Kyushu University, Fukuoka 812-8581, Japan}
\date{ }
\maketitle
\begin{abstract}
Emergence of hierarchies is investigated by Monte
Carlo simulation in a timid society where all individuals are
pacifist. The self-organiztion of hierarchies is shown to occur
in two steps as the population is increased, i.e. there are three
states, one egalitarian and two hierarchical states;the transition
from the egalitarian to the first hierarchical state is
continuous and the transition from the first hierachical state to
the second one is discontinuous.
In the first hierarchical society, all individuals belong to
either middle class or losers and no winners appear. In the
second hierarchical society, many winners emerge and the population
of the middle class is reduced. The hierarchy in the
second hierarchical society is stronger than the hierachy in a
no-preference society studied by Bonabeau et al [ Physica A{\bf 217},
373 (1995)]
\end{abstract}

\section{Introduction}
The emergence of hierarchies is a common phenomenon in societies and animal
groups. In a pioneering work, Bonabeau {\it et al.}\cite{bonabeau} have
shown that a hierarchical society can emerge spontaneously from an
equal society by a simple algorithm of fighting between individuals
diffusing on a square lattice.
On the basis of results of Monte Carlo simulation and an analysis by a mean
field theory, they concluded that subcritical or supercritical bifurcations
exist in the formation of the hierarchical structure as the density of
individuals is varied. 
In their model, each individual is assumed to have some wealth or power which
increases or decreases by winninng or losing in a fight.
The essential processes of the model are diffusion, fighting
and spontaneous relaxation of the wealth.
Various societies can be modelled by specifying each process and
the emergence of the hierarchy depends strongly on the
specifications.\cite{sousa,stauffer}

In this paper, we investigate a variation of the model introduced
by Bonabeau {\it et al.}\cite{bonabeau},
where the diffusion algorithm is modified to
include the effect of the trend of society. Namely, we study the emergence of
hierarchies in a timid society, in which an individual always tries
to avoid fighting and to fight with the weakest among the neighbors
if he/she cannot avoid fighting.
By Monte Carlo simulation, we show that the emergence of the hierarchy
is retarded in the timid society compared to the no-preference society
investigated by Bonabeau {\it et al.} and that the transition
to the hierarchical state occurs in two successive transitions of
a continuous and a discontinuous ones.
Consequently, there exist
three different states in the society, one equal and two hierarchical states.
In the first hierarchical states, we see no winners but losers and
people in the middle class. In the second hierarchical states, 
many winners emerge from the middle class.
We also show that the distribution of wealth in the second hierarchical
state is wider compared to the hierarchical state of the no-preference society.

In Sec. 2, our model is explained in detail.
Results of Monte Carlo simulation are presented in Sec. 3.
In Sec. 4 the characteristics of the hierarchical
states is analyzed in detail. Section 5 is devoted to discussion.

\section{A timid society}
We consider $N$ individuals diffusing on an $L \times L$ square lattice,
where every lattice site accomodates at most one individual.
An individual is to move to one of nearest neighbor sites
according to the following protocol.
When individual $i$ tries to move to a site occupied by $j$,
$i$ and $j$ fight each other. If $i$ wins, $i$ and $j$ exchange their
positions, and if $i$ loses, they keep their original positions.
We associate each individual a quantity which we call power or wealth.
The power increases by unity for every victory and decreases
by unity for every loss.
The probability $Q_{ij}$ that $i$ wins the fight against $j$ is determined
by the difference of their powers $F_i$ and $F_j$ as
\begin{equation}
Q_{ij} = \frac{1}{1 + \exp[\eta(F_j - F_i)]} ,
\label{probability}
\end{equation}
where $\eta$ is introduced as a controlling parameter.
When $\eta = \infty$, the stronger one always wins the fight
and when $\eta = 0$, the winning probability of both ones are equal.
We also assume that the power of individuals relaxes to zero when 
they do not fight, namely power $F_i(t+1)$ at time $t+1$ is given by
$F_i(t)$  through\cite{bonabeau}
\begin{equation}
F_i (t+1) = F_i(t) - \mu \tanh[F_i(t)] .
\end{equation}
Here, the unit of time is defined by one Monte Carlo step
during which every individual is accessed once for move and
$\mu$ represents an additional controlling parameter.
This relaxation rule indicates that people lose their wealth
of a constant amount when their power is large, and
when their power is small, they lose it at a constant fraction,
namely they behave rather miserly.
It also indicates that the negative wealth (debt) can relax to zero
in the similar manner.
Note that this rerlaxation rule is critical to the emergence of
hierarchical society.\cite{sousa}

We characterize the timid society by the preference of
individuals in diffusion. In the timid society, every individual favors
not to fight and thus it moves to a vacant site if it exists. If no vacant
sites exist in the nearest neighbors, then it moves to a site occupied by
an individual whose power is the smallest among the neighbors.
When more than two neighbors have the equal power, then an opponent
is chosen randomly from them.

We characterize the static status of the society by an order parameter $\sigma$
which is defined by\cite{bonabeau, sousa}
\begin{equation}
\sigma^2 = \frac{1}{N}\sum_{i}\left\{
\frac{D_i}{D_i+S_i} -\frac{1}{2}\right\}^2 .
\label{order}
\end{equation}
Here, $N$ is the number of individuals, and
$D_i$ and $S_i$ are the number of fights won and lost, respectively,
by individual $i$. Note that $\sigma = 0$ corresponds to an egalitarian
status and $\sigma = 1/\sqrt{12} \simeq 0.2887$ when the chance for victory
$\frac{\displaystyle D_i}{\displaystyle D_i+S_i}$ is distributed
uniformly in $[0, 1]$. After sufficiently long Monte Carlo simulation,
variation of $\sigma$ is stabilized and one can use it as an order parametrer.

We also monitor the population profile by focusing on the winning probability.
We classify individuals into three groups by the number of fights which
an individual won; winners are individuals who won more than 2/3 of fights
and losers are individuals who won less than 1/3 of fights.
Individuals between these groups are called middle class.

\section{Monte Carlo simulation}
Monte Carlo simulation was performed for $N = 3500$ individuals
on the square lattice
with periodic boundary conditions from $L = 60$ to $L = 180$.
We obtained the order parameter $\sigma^2$ and other quantities for
$10^6$ Monte Carlo steps. 
 
\begin{figure}[thb] 
\begin{center} 
\includegraphics[height=8cm]{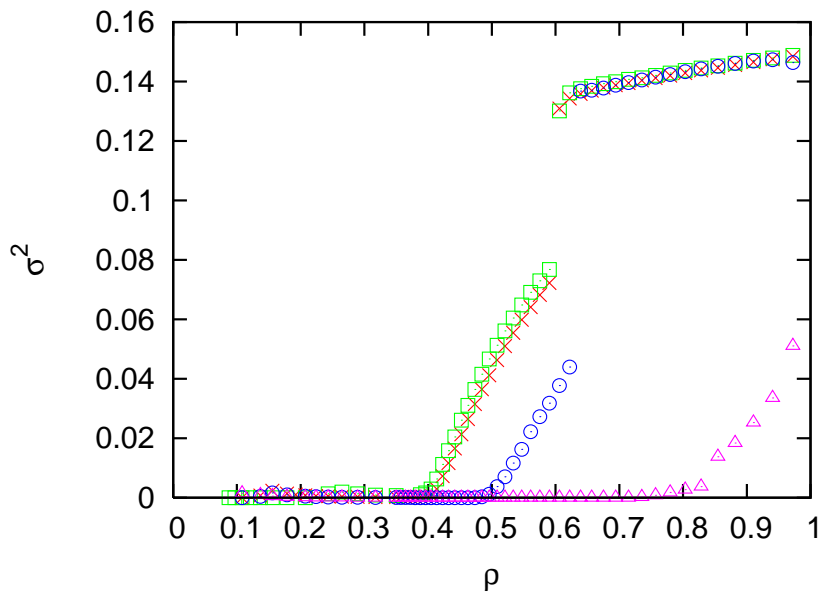} 
\end{center} 
\caption{Order parameter $\sigma ^2$ as a function of $\rho=N/L^2$ with
 $\mu=0.1$, for four different values of $\eta$:
$\eta =$ $50(\Box)$, $0.5(\times)$, $0.05(\bigcirc)$, $0.005(\triangle)$.
Error bars are much smaller than the size of symbols.}
\end{figure} 

Figure 1 shows the dependence of the order parameter $\sigma^2$
on the density $\rho = N/L^2$ for several values of $\eta$, where
$\mu$ is fixed to $\mu = 0.1$.
We can see two clear transitions; one at a lower critical density
$\rho_{C1}$ and the other at a higher critical density $\rho_{C2}$.
The transition at $\rho_{C1}$ is continuous and the transition
at $\rho_{C2}$ is discontinuous.
The dependence of the critical densities $\rho_{C1}$ and $\rho_{C2}$
on parameter $\eta$ is shown in Fig. 2.

\begin{figure}[bht] 
\begin{center} 
\includegraphics[height=8cm]{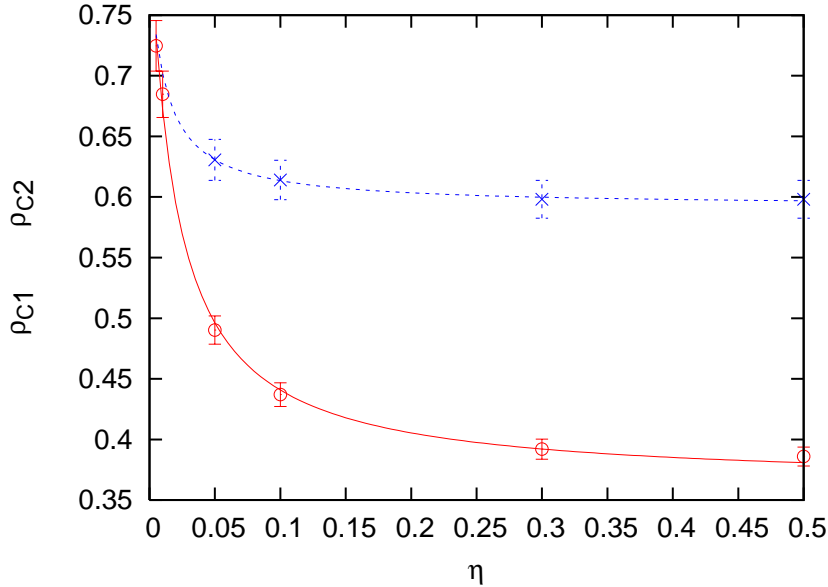} 
\end{center} 
\caption{The dependence of the critical densities $\rho_{C1}$
 (the circles) and $\rho_{C2}$ (the crosses) on parameter $\eta$,
where $\mu=0.1$. The curves are the guide for eyes.}
\end{figure} 

We can identify three states for a given value of $\eta$;
an egalitarian state for $\rho \leq \rho_{C1}$, a hierarchical
society of type I for $\rho_{C1} \leq \rho \leq \rho_{C2}$
and a hierachical society  of type II for $\rho_{C2} \leq \rho \leq 1$.
In the egalitarian society, winners and losers lose their
memory of previous fight before they engage in the next fight, and
thus they changes their status in time.
In the hierarchical state, a winner keeps winning and a loser keeps losing.
We discuss the difference between type I and type II hierarchical society
in the next section.

The results strongly depend on $\mu$. We show the phase
boundary on the $\rho$-$\mu$ plane for $\eta = 0.05$ in Fig. 3.
\begin{figure}[bht] 
\begin{center} 
\includegraphics[height=8cm]{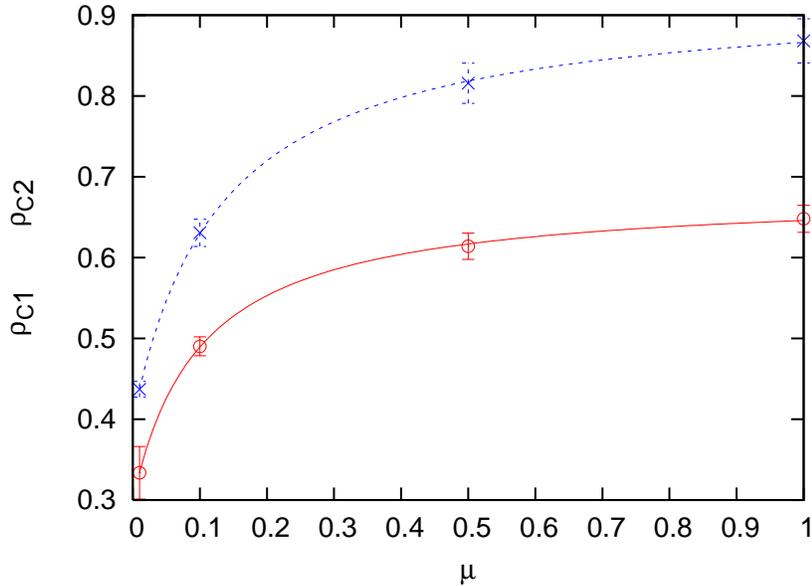} 
\end{center} 
\caption{The dependence of the critical densities $\rho_{C1}$
 (the circles) and $\rho_{C2}$ (the crosses) on parameter $\mu$
for $\eta = 0.05$. The curves are the guide for eyes.}
\end{figure} 

\section{Two hierarchical societies}
In order to investigate the structure of the hierarchical states,
we analyze profile of population.
The dependence of the population of each class is plotted
against the density in Fig. 4.
Rapid changes of the populations signify emergence of different
state of the hierarchical societies.
In the egalitarian state $\rho \leq \rho_{C1}$, 
all individuals belong to the middle class as expected.
In the hierachical society I $\rho_{C1} \leq \rho \leq \rho_{C2}$,
some individuals become losers whose number increases as
the density is increased, but no winners are seen.
In the hierachical state II $\rho \geq \rho_{C2}$,
many winners appear and the population in the middle
class is reduced significantly.
\begin{figure}[thb] 
\begin{center} 
\includegraphics[height=8cm]{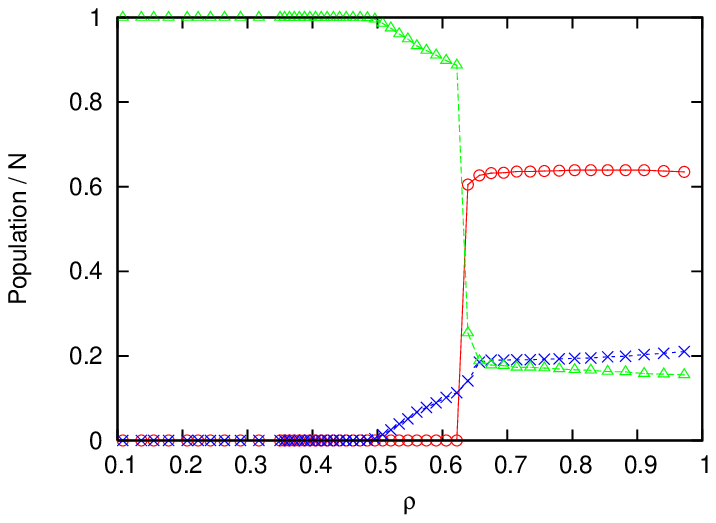} 
\end{center} 
\caption{Dependence of the population in each class on the density
when $\mu = 0.1 $ and $\eta = 0.1$.
}
\end{figure} 

Figures 5 (a), (b) and (c) show the spacial distribution of individuals after
 $10^6$ Monte Carlo steps in the egalitarian, the first hierarchical and the
second hierarchical states, respectively. No specific spatial
inhomogeneity is observed in the timid society.

\begin{figure}[htb] 
\begin{center}
(a) \includegraphics[height=6.3cm]{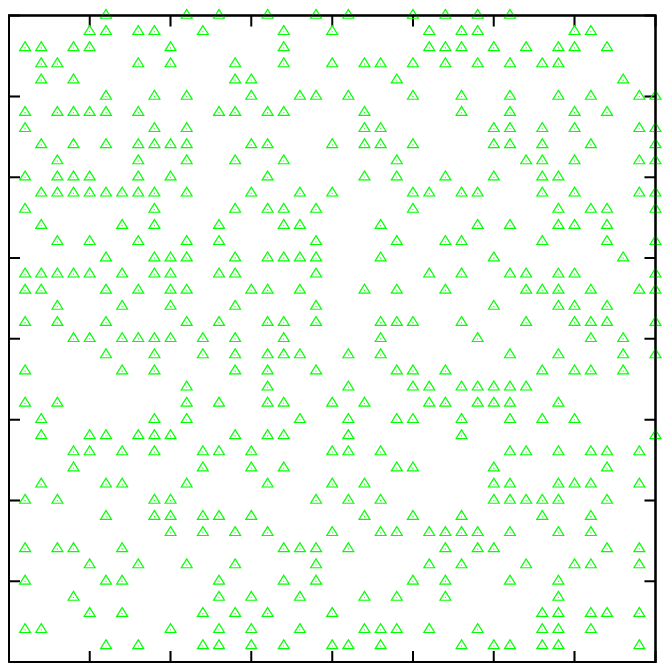} \\
(b) \includegraphics[height=6.3cm]{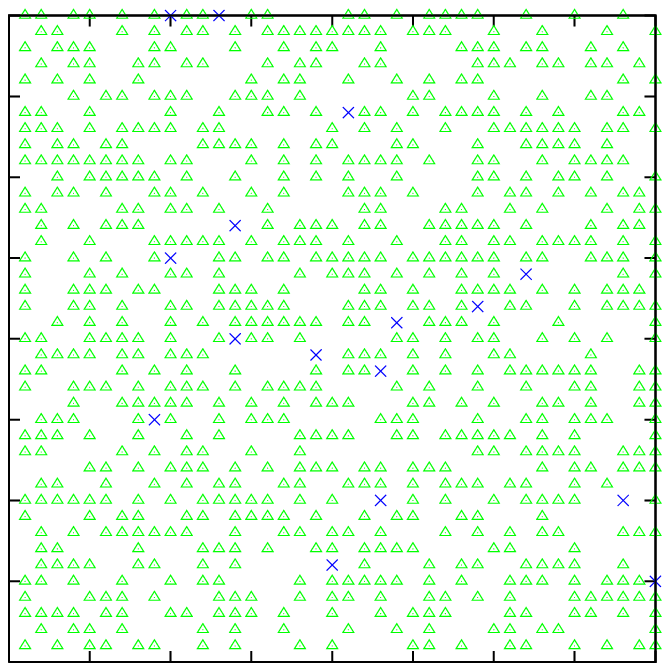} \\
(c) \includegraphics[height=6.3cm]{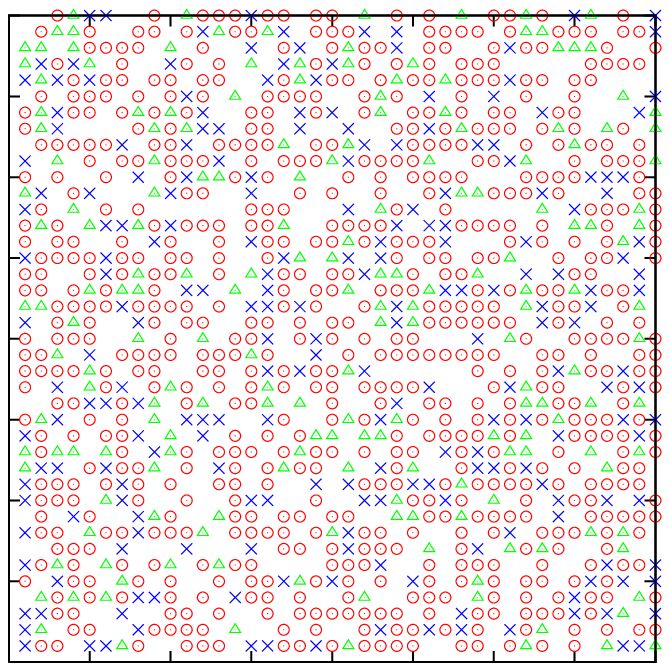}
\end{center}
\caption{Structure in the quilibrium state. (a) the eagalitarian state
$\rho = 0.3$,
(b) the hierarchical socity I $\rho = 0.5$
and (c) the hierarchical socity II $\rho = 0.7$.
the circles are the winner, the traiangles are individuals in the
middle class and the crosses are the loser.}
\end{figure}

In order to see details of the hierarchical structure, we plot the
population as a function of the density and the winning probablity in Fig. 6.
From this plot, we conclude that (1) in the hierarichical state I,
people in the middle class with slightly higher winning probability
increase, but no winners are seen and (2) in the hierarchical state II,
the most of winners have very high winning probability, while people in the
losers and the middle class are distributed in a wide region of the winning
probability, 

\begin{figure}[bht] 
\begin{center} 
\includegraphics[height=8cm]{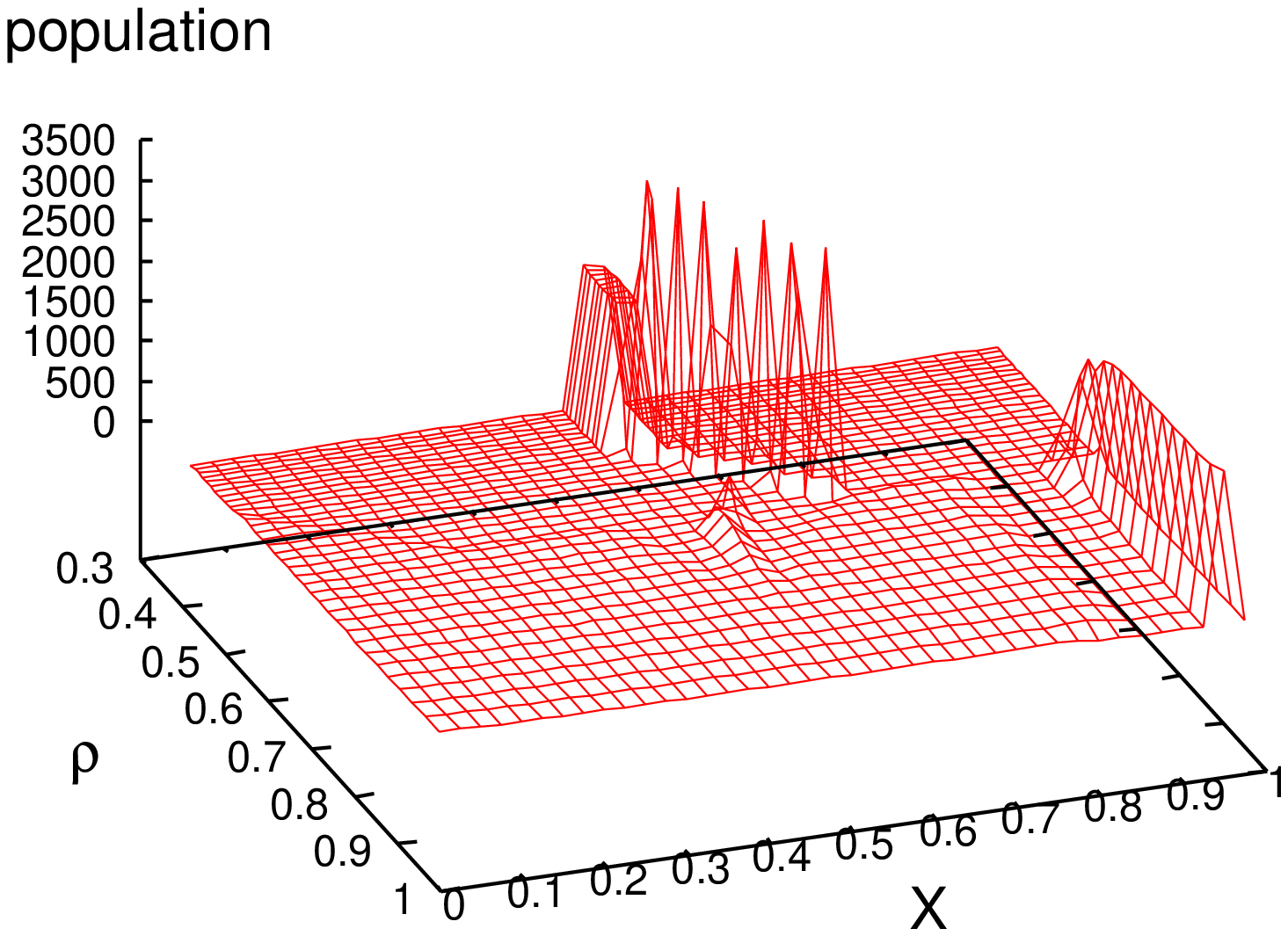}
\end{center} 
\caption{Population as a function of the density and
the winning rate. $X = D_i/(D_i + S_i)$.}
\end{figure}
\section{Discussion}
We have investigated the emergence of self-organized hierarchies
in the timid society.
Our results show that the emergence of the hierarchical state
in the timid society is retarded compared to the no-preference society.
This delay is natural since individuals in the timid society
tend to avoid fighting and thus the wealth is distriruted
more or less evenly among individuals when the population is low.
Furthermore, the emergence of the hierarchical society in the timid
society occurs in two steps, and the first transition is continuous
and the second one is discontinuous.
The strength of the hierarchy in the high density region is stronger
in the timid society compared to the no-preference society.
For the same choice of $\eta = 0.05$ and $\mu = 0.1$, $\sigma^2$
for the former case is twice as large as the latter\cite{bonabeau}.

To understand these behaviors, we first remind the fact that the
the hierarchical society emerges when the power cannot relax before the
subsequet fight. In the timid society, an idividual can avoid fighting
when the density is low, and  thus the ealitarian state is favored
for low densities. In the timid society, weaker individuals have more chance
to be challenged and thus to lose their power, and stronger ones has less
chance to fight and their power stay near zero. This situation corresponds
to the hierarchical state I.
When the density is increased above the upper critical density,
all individuals have more chance to fight and thus stronger individuals become
much stronger.





\end{document}